\documentclass{iau}
\usepackage{graphicx}

\title[Cold and hot gas flows feeding forming ETGs] 
{The role of cold and hot gas flows in feeding early-type galaxy formation}

\author[Peter H. Johansson]   
{Peter H. Johansson$^1$}

\affiliation{$^1$Department of Physics, University of Helsinki, \\ Gustaf H\"allstr\"omin katu 2a,
FI-00014 Helsinki, Finland \\ email: {\tt Peter.Johansson@helsinki.fi} \\[\affilskip]}

\pubyear{2014}
\volume{308}  
\pagerange{xx--xx}
\jname{The Zeldovich Universe: Genesis and Growth of the Cosmic Web}
\editors{R. van de Weygaert, S. Shandarin, E. Saar \& J. Einasto, eds.}
\begin{document}

\maketitle

\begin{abstract}
We study the evolution of the gaseous components in massive simulated galaxies and show that their 
early formation is fuelled by cold, low entropy gas streams. At lower redshifts of $z\lesssim 3$ the simulated
galaxies are massive enough to support stable virial shocks resulting in a transition from cold to hot gas accretion. 
The gas accretion history of early-type galaxies is directly linked to the formation of their stellar component in the
two phased formation scenario, 
in which the central parts of the galaxy assemble rapidly through in situ star formation and the later assembly 
is dominated primarily by minor stellar mergers.

\end{abstract}

\firstsection 
\section{Introduction}

Recently there has been growing evidence both observationally (e.g. \cite[Bezanson et al.  2009]{2009ApJ...697.1290B};
\cite[Toft et al.  2014]{2014ApJ...782...68T}) and 
theoretically (e.g. \cite[Naab et al. 2007]{2007ApJ...658..710N}, \cite[2009]{2009ApJ...699L.178N}) that massive early-type
galaxies assemble in two phases. At high redshifts of $z\sim 3-6$ the central parts of the galaxies assemble
rapidly through compact in situ star formation fuelled by cold, low entropy gas streams penetrating deep within
the galactic halo. At lower redshifts the in situ growth of the stellar component is stifled by the lack of cold
gas and instead the late assembly proceeds predominantly through the accretion of stars formed in subunits 
outside the main galaxy 
(\cite[Oser et al.  2010]{2010ApJ...725.2312O}, \cite[2012]{2012ApJ...744...63O}; \cite[Lackner et al.  2012]{2012MNRAS.425..641L}).

Here we study in more detail the evolution of the gaseous component in simulated early-type galaxies and show that
the two phased formation of the stellar component is driven by a bimodal evolution in the temperatures of the gas
accreting onto the forming galaxies. The results are based on a subsample of four galaxies (A2, C2, E2 and U) extracted from 
\cite[Johansson et al. (2012)]{2012ApJ...754..115J} and simulated at high spatial ($\epsilon_{\star}=0.125-0.25 \ \rm kpc$)
and mass resolution ($m_{\star}\sim 10^{5}-10^{6} M_{\odot}$). All of the simulations were run using the 
Gadget-2 smoothed particle hydrodynamics (SPH) code (\cite[Springel 2005]{2005MNRAS.364.1105S}) and include cooling for a primordial composition, 
star formation and  self-regulating feedback from type II supernovae (\cite[Springel \& Hernquist 2003]{2003MNRAS.339..289S}). 

\section{The gas assembly of early-type galaxies}

\begin{figure}[h]
\begin{center}
 \includegraphics[width=10.8cm]{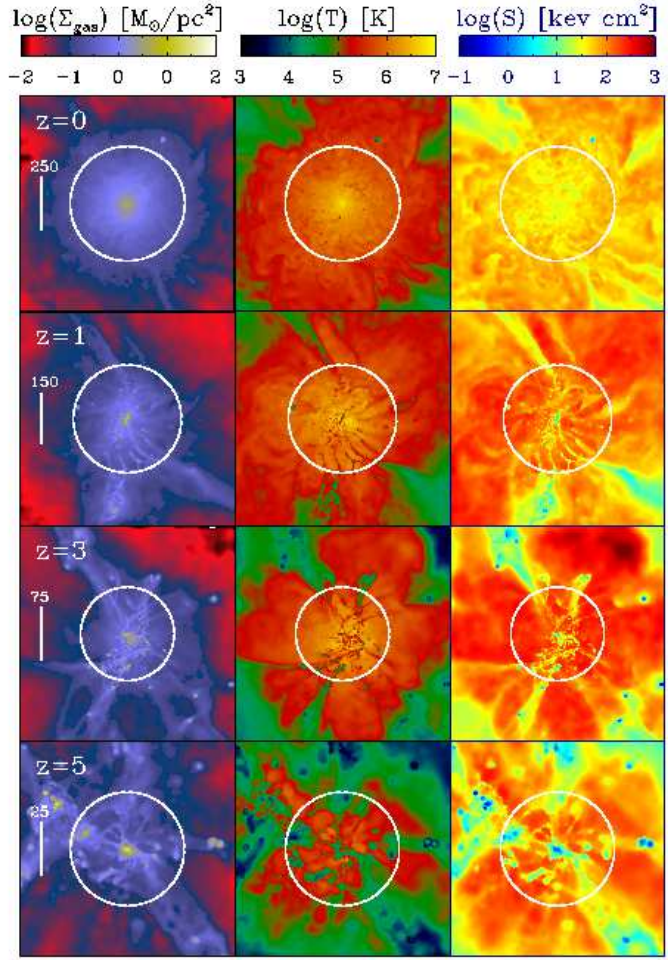} 
\caption{The gas surface density (left panel), the mass-weighted mean gas temperature (middle panel) and mass-weighted
mean gas entropy $(S=kTn^{-2/3}$, right panel) are shown for halo A2 at redshifts of $z=0$, $z=1$, $z=3$ and $z=5$ from top to
bottom. The length scale in kpc is indicated by the white bar and the white circles show the corresponding virial radii.}
   \label{fig1}
\end{center}
\end{figure}

In Fig. \ref{fig1} we depict the evolution of the gas surface densities, together with the mass-weighted temperatures and 
entropies for halo A2, which is representative of our simulation sample. 
At the high redshift of $z=5$ the forming galaxy is found sitting at the intersection
of several gas filaments, which are feeding cold gas directly onto the galaxy as can be seen in the blue filaments in the entropy panel. 
At the lower redshifts of $z=3$ and especially at $z=1$ the halo has grown sufficiently in mass to be able to support strong virial shocks
resulting in a majority of hot gas. However, at both of these redshifts there are still some remaining cold gas filaments
that are able to penetrate into the central gaseous structure and supply some fuel for star formation. In the final snapshot
at $z=0$ almost all of the gas can be found in a hot diffuse state, with virtually no remaining cold clumpy gas.

\section{The temperature structure of the gas}

\begin{figure}[t]
\begin{center}
 \includegraphics[width=13.0cm]{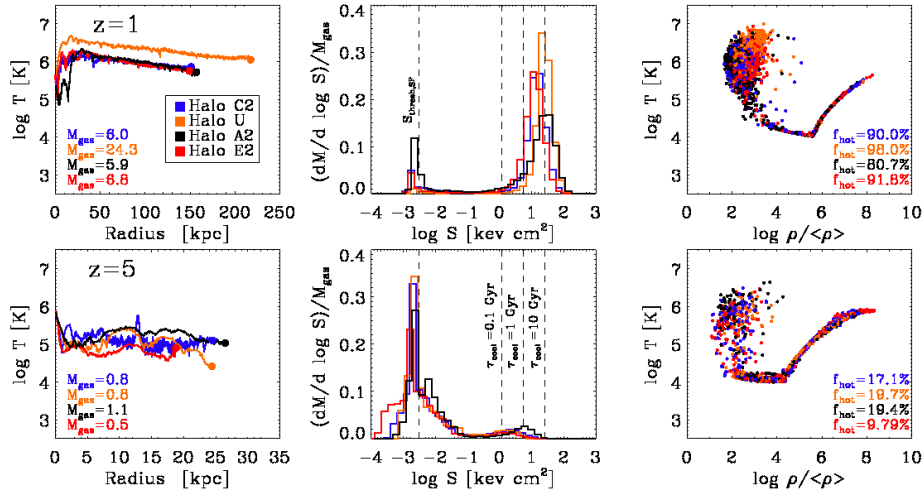} 
 \caption{The temperature profile (left panel), the entropy distribution (middle panel) and the phase-space diagram (right panel)
         for gas within the virial radius for four simulated galaxies at redshifts of $z=1$ (top panels) and
         $z=5$ (bottom panels). The gas mass $(M_{\rm gas})$ in units of $10^{10} M_{\odot}$ together with the fraction $(f_{\rm hot})$ of hot diffuse
         gas is also indicated.} 
   \label{fig2}
\end{center}
\end{figure}

A more detailed view of the evolution of the gas properties of the galaxies can be gained from Fig. \ref{fig2}, where we
plot the temperature profiles, entropy distributions and phase-space diagrams for our simulation sample of four galaxies
at redshifts of $z=5$ and $z=1$. The mean temperature of the gas is increasing in all of the simulations from $T\sim 10^{5} \ \rm K$ at $z=5$ 
to temperatures in excess of a million degrees $(T\sim 10^{6} \ \rm K)$ by $z=1$. 

In the middle panel the gas entropy distributions defined as $S=kTn^{-2/3}$ are shown, where $k$ is the Boltzmann constant, $T$ is the 
temperature and $n$ is the number density. The dashed vertical lines indicate the threshold entropy for star formation (neutral gas
at $T\sim 10^{4} \ \rm K$) and the entropy values corresponding to the 
minimum cooling times of 0.1, 1 and 10 Gyr. The entropy distributions are bimodal, at high redshifts the majority of the gas is cold, high-density,
star-forming gas that forms a low-entropy peak. By redshift $z=1$ most of the cold gas has formed stars and the remaining gas is dilute shock-heated
gas with long cooling times forming a high-entropy peak. Correspondingly the fraction of hot gas (defined as $T> 2.5 \times 10^{5} \ \rm K$ and
$n < n_{\rm thresh,SF}=0.205 \ \rm cm^{-3}$) increases from $f_{\rm hot}\lesssim 20\%$ at $z=5$ to $f_{\rm hot}\gtrsim 90\%$ by $z=1$. At $z=5$ 
star-forming gas can clearly be seen in the phase-diagram plot as most of the gas has a density of $\rho>\rho_{\rm thresh,SF}$ and can be found on an 
equilibrium curve in the $\rho-T$ plane dictated by the self-regulated feedback model (\cite[Springel \& Hernquist 2003]{2003MNRAS.339..289S}) 
employed in this study. At $z=1$, although the majority
of the gas is in the hot component, some residual star-forming gas can also be found. This is in contrast with the no feedback simulations of 
\cite[Johansson et al. (2009)]{2009ApJ...697L..38J} in which virtually no cold star-forming gas remained at redshifts below $z\lesssim 2$.

\section{Cold and hot gas flows}

\begin{figure}[t]
\begin{center}
 \includegraphics[width=9.5cm]{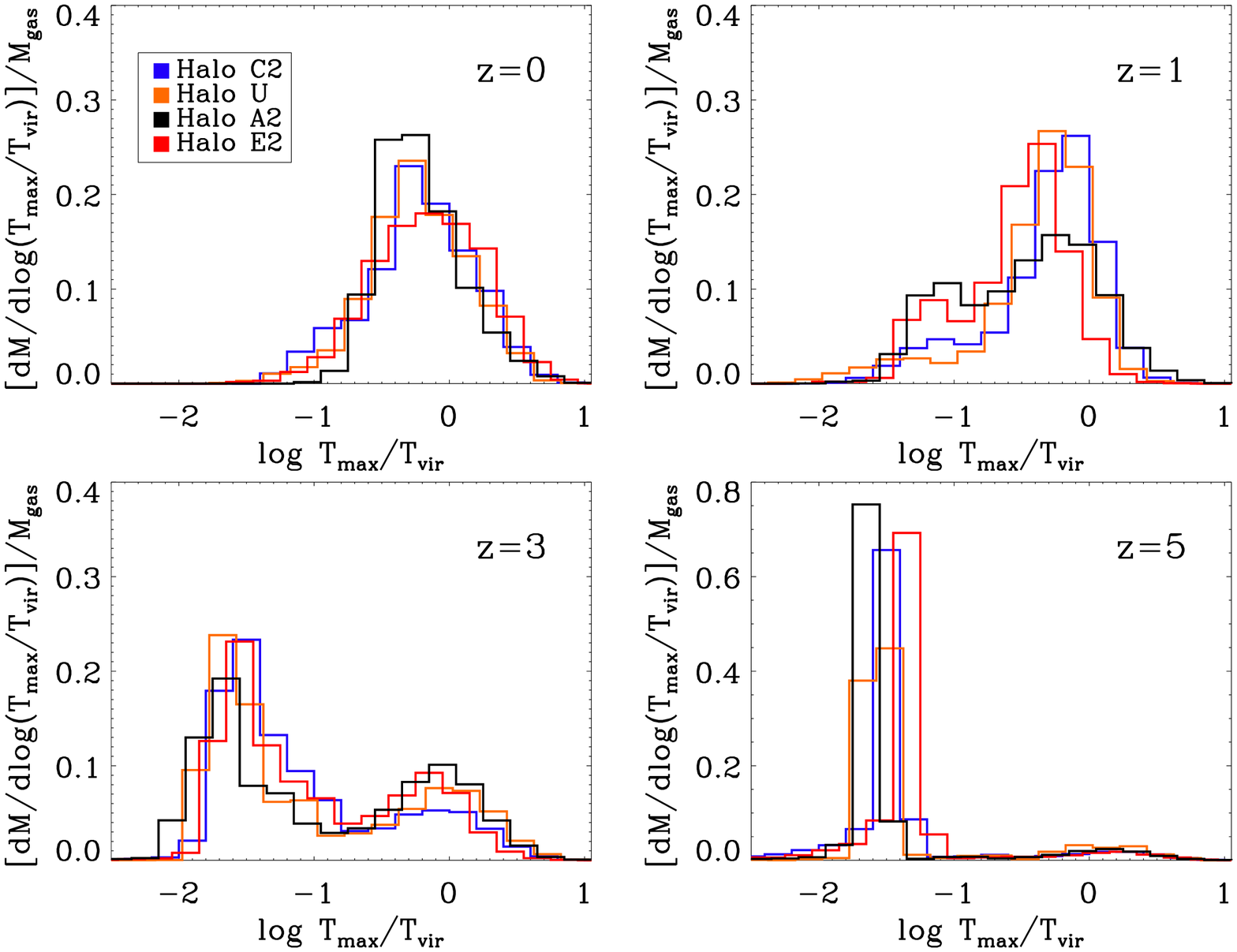} 
 \caption{The distribution of the maximum gas temperatures normalized to the virial temperature of gas accreting onto the four simulated 
          galaxies as a function of redshift. At high redshifts the gas predominantly flows in cold, whereas at lower redshifts the gas is
          accreted in a hot mode.}
\label{fig3}
\end{center}
\end{figure}

In Fig. \ref{fig3} we study following \cite[Kere{\v s} et al. (2005)]{2005MNRAS.363....2K} the thermal properties 
of the accreted gas particles prior to their
accretion onto the haloes.  
The temperature evolution of the accreted gas particles is traced back
through the simulation and the maximum temperature is recorded excluding the snapshots when the gas particles were star-forming as
in these cases the high temperature was due to supernova feedback. 

From Fig. \ref{fig3} we see that at $z=5$ almost all of the gas is accreted with a temperature that is below one tenth of the
virial temperature, which is typically a few times $10^{5} \ \rm K$ at $z\sim 5$. This indicates that the gas is accreted cold at
a typical temperature of a few times $10^{4} \ \rm K$ at this redshift. At $z=3$ the distribution is more bimodal with the majority
of accreted gas still being cold, however a substantial component of hot gas being accreted at $T\sim T_{\rm vir}$ can also be seen. The
transition from predominantly cold accretion to hot accretion occurs at $z\sim 2-3$ when the galaxies reach masses of 
$M_{\rm halo}=5\times 10^{11}-10^{12} M_{\odot}$ in good agreement with the predictions of \cite[Dekel \& Birnboim (2006)]{2006MNRAS.368....2D}. 
At lower redshifts the haloes
are sufficiently massive to support stable virial shocks and the vast majority of the gas is accreted in the hot phase. 

The transition from a gas accretion model based on cold gas flows at $z\gtrsim 3$ to a hot accretion mode at lower redshifts $(z\lesssim 3)$ 
is directly connected to the two phased picture of early-type galaxy formation. In this picture the early formation is dominated by in situ star formation
fuelled by cold gas flows, whereas the later assemble history is dominated by stellar accretion primarily through minor mergers as the source
of internal cold gas is exhausted by the transition from cold gas accretion to a hot accretion mode.

\end{document}